\documentclass[aps,pra,twocolumn,groupedaddress,showpacs]{revtex4}
\usepackage{amssymb}
\usepackage{amsfonts}
\usepackage{amsmath}

\def\Det{{\bf Det}}

\def\<{\langle}
\def\>{\rangle}
\def\H{{\cal H}}

\def\C{{\mathbb C\, }}

\def\SG{{\mathfrak S}}
\def\PP{{\mathbb P}}
\def\xx{{\bf x}}
\def\yy{{\bf y}}
\def\zz{{\bf z}}

\begin{document}

\title{The polynomial invariants of four qubits}

\author{Jean-Gabriel Luque}
\email{luque@univ-mlv.fr}
\author{Jean-Yves Thibon}
\email{jyt@univ-mlv.fr}
\affiliation{Institut Gaspard Monge, Universit\'e de Marne-la-Vall\'ee\\
77454 Marne-la-Vall\'ee cedex, France}

\date{December 14, 2002}

\begin{abstract}
We describe explicitly the algebra of polynomial functions on the Hilbert
space of four qubit states which are invariant under the
SLOCC group $SL(2,\C)^{4}$. From this description, we obtain
a closed formula for the hyperdeterminant in terms of low degree invariants.
\end{abstract}


\maketitle

\section{\label{intro}Introduction}

Various classifications of states with up to four qubits have
been recently proposed, with the aim of understanding the different
ways in which multipartite systems can be 
entangled \cite{Dur,Bry, Bry2,Ver,Mi}.
However, one cannot expect that such classifications will
be worked out for an arbitrary number $k$ of qubits, and there
is a need for a coarser classification scheme which would
be computable for general $k$. In  \cite{Klyachko},
Klyachko proposed to assimilate entanglement with the
notion of {\it semi-stability} of geometric invariant theory.
In this context, a semi-stable state is one which can be separated
from 0 by a polynomial invariant
of $SL(2,\C)^{k}$, the point in the geometric approach
being that explicit knowledge of the invariants is in principle not
necessary to check this property.

In this paper, 
we construct  a complete set of algebraic
invariants of 4-qubit states. This allows us to identify
the semi-stable states in the classification of Verstraete
et al. \cite{Ver}, and to obtain a simple closed form for 
the hyperdeterminant.

Let $V=\C^2$ be the local Hilbert space of a spin $\frac12$ particle, 
and  $\H=V^{\otimes 4}$ be the state space of four
particles, regarded as the natural
representation of the 
group $G=SL(2,\C)^4$,
known in the context of quantum information
theory (QIT) as the group of reversible stochastic local
quantum operations assisted by classical communication (SLOCC) \cite{Ben,Dur}.

If $|j\>$,
$j=0,1$ is any basis of $V$, a state $|\Psi\>$ can be written as
\begin{equation}
|\Psi\>=\sum_{i,j,k,l=0}^1 A_{ijkl} |i\>\otimes |j\>\otimes |k\>\otimes |l\>
\end{equation}
and the question
of which normal form can be achieved for $|\Psi\>$
by varying independently the  bases of the four copies of $V$ has
been solved only recently \cite{Ver} , although the case of three-qubit states
is classical and relatively  simple \footnote{
In the physics literature, the solution appears in \cite{Dur},
but an equivalent problem had been solved in classical invariant theory
since at least 1881 \cite{LeP,Sw}. Note also that a problem
equivalent to the classification of four-qubit states
had been studied by Segre in 1922 \cite{Segre}, but he only
obtained a partial classification.}.

In the following, we give a complete 
\footnote{
We note that a combinatorial method for computing invariants of
fourth-rank tensors is proposed in \cite{TaP}. The point in the present
paper is that we can prove that our system of invariants is complete.}   
description of the polynomial functions
$f(A_{ijkl})$ which are invariant under 
the SLOCC group $SL(2,\C)^4$.
This  amounts to the construction of a moduli space for four qubit states.
Our strategy is to  find  first the Hilbert series of the algebra of
invariants ${\cal J}$. Next, we construct by classical methods four invariants
of the required degrees. The knowledge of the Hilbert series reduces then
the proof of  algebraic independence and completeness to simple verifications.
The values of the invariants on the orbits of \cite{Ver} are tabulated
in the Appendix.

\section{\label{hilbert}The Hilbert series}

Let ${\cal J}_d$ be the space of $G$-invariant homogeneous polynomial
functions of degree $d$ in the variables $A_{ijkl}$. Using some elementary
representation theory, it is not difficult to show that  ${\cal J}_d$ is
zero for $d$ odd, and that for $d=2m$ even, the dimension of  ${\cal J}_d$
is equal to the multiplicity of the trivial character of the
symmetric group $\SG_{2m}$ in the fourth power $(\chi^{mm})^4$ of
its irreducible character corresponding to the partition $[m,m]$.
This is the same as the scalar product $\<(\chi^{mm})^2|(\chi^{mm})^2\>$,
which can be evaluated by means of the formulas of \cite{Th91,STW}
giving the decomposition into irreducible characters of any product
$\chi^\lambda\chi^\mu$ when $\lambda$ and $\mu$ have at most two parts.
This yields the Hilbert series of ${\cal J}=\bigoplus_d{\cal J}_d$
in the form \footnote{
This is a tedious method. With the help of computer algebra system,
the Hilbert series is more straightforwardly obtained by a residue calculation.
Similar computations (for $SU(2)$ and $U(2)$ invariants) appear
in \cite{GBRM}.}
\begin{equation}
\sum_{d\ge 0}\dim {\cal J}_d \, t^d=
\frac{1}{(1-t^2)(1-t^4)^2(1-t^6)} \,.
\end{equation}
This formula shows that Conjecture 2.6.5.3 of \cite{Klyachko}
cannot be correct, since it predicts that the hyperdeterminant,
which is of degree 24, should be one of the generators.
Actually, the algebra of invariants is  free on generators
of degrees $2,4,4,6$, as suggested by the Hilbert series.

\section{\label{invariants}A fundamental set of invariants}

Indeed, it is possible to construct invariants of the
required degrees  and to check that
they are algebraically independent. To reduce the size
of the expressions, we shall write the components of $|\Psi\>$ as
\begin{equation}
A_{ijkl}= a_r\,,\quad r=0,\ldots,15\,,
\end{equation}
where $r$ is the integer whose binary expression is $ijkl$, that is,
$r=8i+4j+2k+l$. We shall consider them as the coefficients of
a quadrilinear form
\begin{equation}
A(\xx,\yy,\zz,{\bf t})=\sum_{i,j,k,l=0}^1A_{ijkl}x_iy_jz_kt_l  
\end{equation}                                                                              
on $V\times V\times V\times V$. Such a form is known to have
an invariant $H$ of degree 2,  which is also one of the hyperdeterminants
introduced by Cayley \footnote{Cayley actually considered several different
notions under the same generic name, see \cite{LT}.}.
It is given by
\begin{eqnarray}
H= &a_0a_{15} -a_1a_{14}-a_2a_{13}+a_3a_{12}\nonumber\\
&-a_4a_{11}+a_5a_{10}+a_6a_9-a_7a_8\,,
\end{eqnarray}
and the two independent invariants of degree 4 are any
two of the 3 determinants which can be formed by interpreting
$A$ as a linear map $\C^4\rightarrow \C^4$ (see \cite{Segre})
\begin{eqnarray}
L=\left|
\begin{array}{llll}
a_0 & a_4 & a_8 & a_{12} \\
a_1 & a_5 & a_9 & a_{13} \\
a_2 & a_6 & a_{10} & a_{14} \\
a_3 & a_7 & a_{11} & a_{15}
\end{array}
\right|
\\
M=\left|
\begin{array}{llll}
a_0 & a_8 & a_2 & a_{10} \\
a_1 & a_9 & a_3 & a_{11} \\
a_4 & a_{12} & a_{6} & a_{14} \\
a_5 & a_{13} & a_{7} & a_{15}
\end{array}
\right|
\\      
N=\left|
\begin{array}{llll}
a_0 & a_1 & a_8 & a_{9} \\
a_2 & a_3 & a_{10} & a_{11} \\
a_4 & a_5 & a_{12} & a_{13} \\
a_6 & a_7 & a_{14} & a_{15}
\end{array}
\right|
\\      
\end{eqnarray}
One has the relation
\begin{equation}
L+M+N=0\,,
\end{equation}
but it is easily checked that any two of them are linearly independent, 
and also that $H^2$ cannot be expressed as a linear combination of them.

To construct a sextic invariant algebraically independent
from the previous ones, 
we shall apply the methods of classical invariant theory,
and
first find some covariants, that is, homogeneous 
$G$-invariant polynomials
in the form coefficients $A_{ijkl}$ and in the original
variables (see, e.g., \cite{Olver} for a modern presentation).
The dimension of the space ${\cal C}_{d,k_1,k_2,k_3,k_4}$
of covariants which are of degree $d$ in $A$, $k_1$ in $\xx$, and
so on, is equal to the multiplicity of the trivial character of
$\SG_d$ in the product
\begin{equation}
\chi^{l_1+k_1,l_1}\chi^{l_2+k_2,l_2} \chi^{l_3+k_3,l_3} \chi^{l_4+k_4,l_4} 
\end{equation}
where $d=2l_i+k_i$ for all $i$.
This can still be evaluated from the knowledge of the products
$\chi^\lambda\chi^\mu$ of two-part characters, 
and one can see that $A$ has
six covariants of degree 2, which are biquadratic forms in
all possible pairs of variables. Such covariants are easily
constructed, these are the determinants of order 2 
of the partial derivatives of $A$ with respect to the complementary
pair of variables, e.g.,
\begin{equation}
b_{xy}(\xx,\yy)=\det\left(\frac{\partial^2A}{\partial z_i\partial t_j}
\right)\,.
\end{equation}
Each of these biquadratic forms can be interpreted as a bilinear
form on the three-dimensional space $S^2(\C^2)$, and one
can define $3\times 3$ matrices by
\begin{equation}
b_{xy}(\xx,\yy)=[x_0^2, x_0x_1, x_1^2] B_{xy}
\left[\begin{array}{c}y_0^2\\ y_0y_1\\ y_1^2\end{array}\right]
\end{equation}
and similarly for the other pairs. Let, for any pair of vector
variables $({\bf u},{\bf v})$,
\begin{equation}
D_{uv}=\det( B_{uv} ) \,.
\end{equation}
These determinants are sextic invariants of $A$. Since
the space of sextic invariants is four-dimensional, they must be
linearly dependent. In fact,
\begin{equation}
D_{xy}=D_{zt}\,,\ D_{xz}=D_{yt}\ {\rm and}\ D_{xt}=D_{yz}
\end{equation}
but $D_{xy}$, $D_{xz}$ and $D_{xt}$ are linearly independent.
One can check that
\begin{eqnarray}
H L &= D_{xz}-D_{xt}\,\\
H M &= D_{xt}-D_{xy}\,\\
H N &= D_{xy}-D_{xz}\,
\end{eqnarray}
and that $H^3$ is not in the subspace spanned by the $D_{uv}$. 
The above results, together with the knowledge of the
Hilbert series, allows now to prove that the
algebra of invariants is free, and that any of the $D_{uv}$'s can
be taken as the generator of degree 6. Indeed, it is sufficient to
check that the Jacobian matrix of the choosen generators has
rank 4 (this is  easily done using the specialization
$G_{abcd}$ of the Appendix).

We will use
in the sequel
\begin{equation}
{\cal J}=\C[H,L,M,D_{xt}]\,.
\end{equation}

\section{\label{hypdet}The hyperdeterminant
in terms of  the fundamental invariants}

Here, according to the general formula of \cite{GKZ},
the Cayley hyperdeterminant (in the sense of \cite{GKZ,Mi})
is of degree 24. It must therefore
admit an expression in terms of the fundamental invariants, whose explicitation
is an interesting question. To answer it, we shall need again the covariants
$b_{uv}$. Let us use, for example, $b_{xt}$. 
We can consider $A$ as a trilinear form $T$ in $\xx,\yy,\zz$, the fourth
variable
${\bf t}$ being treated as a parameter. The Cayley hyperdeterminant
$\Det(T)$ of this trilinear form is homogeneous of degree 4 in its
coefficients, which are themselves linear forms in ${\bf t}$.
Hence, $R({\bf t})=\Det(T)$ is  a binary quartic in $t_0,t_1$, and we can form
its discriminant $\Delta$, which will be an invariant of $A$.
According to Schl\"afli \cite{Schl} (see also \cite{GKZ,Mi}),  
in this case,
$\Delta$ is  equal to $\Det(A)$.

It follows from the well-understood invariant
theory of  binary trilinear forms \cite{LeP,Sw}
that  $R({\bf t})$ is equal to the discriminant
of the quadratic form in $\xx$
\begin{equation}
Q_t(\xx)=b_{xt}(\xx,{\bf t})\,,
\end{equation}
that is
\begin{equation}
R({\bf t})=\det\left(\frac{\partial^2b_{xt}}{\partial x_i\partial x_j}\right)\,.
\end{equation}
Let
\begin{equation}
R({\bf t})=
c_0t_0^4+4c_1 t_0^3t_1+6c_2 t_0^2 t_1^2+4t_0 t_1^3+c_4 t_1^4\,.
\end{equation}
It is well-known that the algebra of invariants of the binary quartic
is free over the two generators
\begin{eqnarray}
S &=&c_0c_4-4c_1c_3+3c_2^2\,,\\
T &=& c_0c_2c_4-c_0c_3^2+2c_1c_2c_3-c_1^2c_4-c_2^3\,,
\end{eqnarray} 
and that its discriminant is given by
\begin{equation} 
\Delta=S^3-27\, T^2\,.
\end{equation}
In the classical language, $S$ is the apolar of $R$ with itself,
and $T$ is its catalecticant (see \cite{Olver}).

The invariants $S$ and $T$ of $R$ being obviously invariants
of $A$, the problem of expressing $\Det(A)$ is terms of
the fundamental invariants of $A$ is reduced to the one of finding
the expressions of $S$ and $T$ \footnote{The values of $S$ and
$T$ may depend on our choices of the pair of variables
$(x,t)$ and then on the choice of eliminating $x$, but
the combination $S^3-27T^2$ will not depend on these choices.}.

With the help of a computer algebra system, we obtain the values
\begin{eqnarray}
S& = & \frac{1}{12} H^4-\frac23H^2L+\frac23H^2M-2H D_{xt}\nonumber\\
&    & +\frac43(L^2+LM+M^2)
\end{eqnarray}
and
\begin{eqnarray} 
T&= & \frac{1}{216} H^6-\frac{1}{18}H^4(L-M)-\frac16H^3D_{xt}
\nonumber\\
&&+\frac19 H^2(2L^2-LM+2M^2)+\frac23H(L-M)D_{xt}\nonumber\\
&&-\frac{8}{27}(L^3-M^3)-\frac49 LM(L-M)+D_{xt}^2\,.
\end{eqnarray}
Setting $D=D_{xt}$, $U=H^2-4(L-M)$ and $V=12(HD-2LM)$, these expressions
can be recast into the more elegant form
\begin{eqnarray}
12S&= & U^2-2V\,,\\
216T &=& U^3-3UV+216D^2\,.
\end{eqnarray}
This suggests that 
$U$ and $V$ might have a geometric meaning.
Actually, similar expressions occur in the course of
Schl\"afli's calculations \cite{Schl}. He does not mention
their invariant theoretic meaning, however, and prefers
to end up with an expression of $\Delta$ as a polynomial
in  $H$, $W=D_{xy}+D_{xz}+D_{xt}$, 
$\Sigma=L^2+M^2+N^2$ and $\Pi=(L-M)(M-N)(N-L)$, which
are invariant under permutations of the indices $ijkl$.

\section{\label{concl}Conclusion}

A fundamental issue in QIT is the understanding
of entanglement. However, as pointed out in \cite{Klyachko}, there is no
universal agreement on the precise definition of entanglement
and on what should be its proper measure.  It is apparently this lacune
which motivated recent attempts to obtain a complete classification
of $k$-qubit states under the SLOCC group $G$ \cite{Ben,Dur,Mi}. 
 
Some familiarity with classical invariant theory leaves little
hope that such a classification can be achieved in general. If we
compare with the somewhat easier
classical problem of binary forms, which, in physical
terms, amounts to  the classification of single spin $s$ states
under $SL(2,\C)$, a complete solution is known only up to spin $s=4$
(with still some  unsolved questions  in the case $s=7/2$), and 
most experts agree that the other cases will remain out of reach.

So, it is unlikely  that the complete SLOCC classification
of $k$-qubit states will ever be found for $k\ge 8$, and it is
probable that formidable computational difficulties will arise well
before this value
\footnote{For 5 qubits, the generic orbits depend
on 16 free parameters, and for eight qubits, the moduli space would be
of dimension 231.}. 
Actually, the orbit structure is still
completely unknown  for $k>4$. 

Now, if we adopt the definition of entanglement proposed in \cite{Klyachko},
that is, to identify  entangled states with the semi-stable
vectors of geometric invariant theory,
the main result of the present paper can be interpreted as a numerical
criterion of  entanglement for 4-qubit states. Indeed,
a semi-stable state is by definition a state which can be separated
from 0 by some invariant polynomial. Thus, according to \cite{Klyachko},
an entangled
4-qubit state would be one for which at least one of the four polynomials
$H,L,M,D$ takes a nonzero value. As we shall see below, this definition
needs to be improved. However, it is plausible 
that refined entanglement
measures for four-qubit states might be built
from the absolute values of these invariants.
These would be natural generalizations of the concurrence
$C$ and the 3-tangle $\tau$ in the two and three qubit cases,
which are proportional repectively to the absolute values of the
determinant and of the hyperdeterminant \cite{Mi}, 
the only polynomial invariants in these cases.

From a geometric point of view, our results show that
the moduli space of entangled states is the weighted projective
space $\PP(1,2,2,3)$, which can be embedded as a rational
threefold in 13-dimensional projective space.
Of course, the approach to semi-stability
and moduli spaces by explicit construction of the
polynomial invariants has  its limits,and it is 
unlikely that this can be done for more that 5 qubits. 

Note also that the notion of semi-stability can be used only to
characterize some generic kind of entanglement. 
Indeed, even in the three-qubit cases,
the so-called $W$-state $\frac{1}{\sqrt{3}}(|001\>+|010\>+|100\>)$ is
not semi-stable (its hyperdeterminant is zero), although it should certainly
be considered as entangled
(even in a strong sense, according to \cite{Dur}). 
The natural candidates for constructing
further measures of entanglement appropriate to such states are the
covariants of classical invariant theory, which are
completely known in the three-qubit case \cite{Sw}.

We expect to be able to describe in a forthcoming
paper the  algebra of covariants in the 4-qubit case,
which would not only reproduce the complete classification of
the orbits, but also to give the equations of their closures,
which are algebraic varieties, and provide new insights
about entanglement
measures for unstable states.  Another case whichs seems to
be readily tractable is the case of three spin 1 states.
The geometric classification of the orbits was known by 1938 \cite{TC},
and numerical calculation of the Hilbert series of invariants
up to degree 108 indicates that 
it should be
\begin{equation}
h(t)=\frac{1}{(1-t^6)(1-t^9)(1-t^{12})}\,.
\end{equation}
Since independent invariants $I_6,I_9,I_{12}$ of the
appropriate degrees are known \cite{Chan}, one can consider
that the SLOCC classification
and the semi-stability problems are essentially solved in this case.

Other cases of less practical importance, such as those
including two spin $\frac12$ particles and one
particle of spin $s\ge 1$, are  easily solved. For $s=1$, there
is only one invariant of degree 6, the hyperdeterminant. For
$s=3/2$, the hyperdeterminant is identically zero, but there
is still one invariant of degree 4, which is the only determinant
that can be formed by displaying the components of $|\Psi\>$ in
a $4\times 4$ matrix. Finally, for $s\ge 2$, there are no
invariant polynomials.

\begin{acknowledgments}
We would like to thank A. Klyachko and A. Miyake for their comments on
preliminary versions of this text, and in particular M. Grassl for pointing
out a gap in our original argument, together with a way of filling it.
\end{acknowledgments}

\appendix

\section{\label{app}Application to the classification of Verstraete et al.}

To conclude, let us discuss the semi-stability of the orbits
obtained in \cite{Ver} (see this reference for notation).
For the family $G_{abcd}$, the values of the fundamental invariants
are \footnote{It is interesting to observe that these polynomials
can be regarded as a fundamental system of invariants for the
standard  action of the Weyl group $D_4$ on the four-dimensional
vector space spanned by $a,b,c,d$. }
\begin{eqnarray}
H &=& \frac12(a^2+b^2+c^2+d^2)\,\\
L&=& abcd\,,\\
M&=& 
{\scriptstyle
\left[\left( \frac{c-d}{2}\right)^2 - \left( \frac{a-b}{2}\right)^2\right]
\left[\left( \frac{a+b}{2}\right)^2 - \left( \frac{c+d}{2}\right)^2\right]  
}\,\\
D&=& -\frac14 (ad-bc)(ac-bd)(ab-cd)
\end{eqnarray}
and the hyperdeterminant is
$\frac{1}{256}V(a^2,b^2,c^2,d^2)^2$, where $V$ denotes the Vandermonde
determinant.
For $L_{abc_2}$,
\begin{eqnarray}
H &=&\frac12(a^2+b^2+2c^2)\,\\
L&=&abc^2\,\\
M&=&
{\scriptstyle
\left[c^2-\left( \frac{a+b}{2}\right)^2\right] \left( \frac{a-b}{2}\right)^2}\, \\
D&=&-\frac14 c^2(a-b)^2(ab-c^2)\,,
\end{eqnarray}
and $\Delta=0$.
For $L_{a_2b_2}$,
\begin{eqnarray}
H &=&a^2+b^2\,\\
L&=&a^2b^2\,\\
M&=&0\,\\
D&=&0\,
\end{eqnarray} 
and  $\Delta=0$.
For $L_{ab_3}$,
\begin{eqnarray}
H &=& \frac12(3a^2+b^2)\,\\
L&=& a^3b\,\\
M&=&
{\scriptstyle 
\left[a^2-\left( \frac{a+b}{2}\right)^2\right]\left( \frac{a-b}{2}\right)^2
}\,\\
D&=&\frac14a^3(a-b)^3\,
\end{eqnarray}
and  $\Delta=0$.
For $L_{a_4}$,
\begin{eqnarray}
H &=&2a^2\,\\
L&=& a^4\,\\
M&=&0\,\\
D&=&0\,
\end{eqnarray}
and $\Delta=0$. 
For $L_{a_20_{3\oplus\bar 1}}$, there is still one nonzero invariant
\begin{equation}
H=a^2\,,
\end{equation}
so that all the above 6 families of orbits are semi-stable, whilst the remaining
three are unstable.

\bibliography{qbits5.bib}

\begin{thebibliography}{20}
\expandafter\ifx\csname natexlab\endcsname\relax\def\natexlab#1{#1}\fi
\expandafter\ifx\csname bibnamefont\endcsname\relax
  \def\bibnamefont#1{#1}\fi
\expandafter\ifx\csname bibfnamefont\endcsname\relax
  \def\bibfnamefont#1{#1}\fi
\expandafter\ifx\csname citenamefont\endcsname\relax
  \def\citenamefont#1{#1}\fi
\expandafter\ifx\csname url\endcsname\relax
  \def\url#1{\texttt{#1}}\fi
\expandafter\ifx\csname urlprefix\endcsname\relax\def\urlprefix{URL }\fi
\providecommand{\bibinfo}[2]{#2}
\providecommand{\eprint}[2][]{\url{#2}}

\bibitem[{\citenamefont{D{\"u}r et~al.}(2000)\citenamefont{D{\"u}r, Vidal, and
  Cirac}}]{Dur}
\bibinfo{author}{\bibfnamefont{W.}~\bibnamefont{D{\"u}r}},
  \bibinfo{author}{\bibfnamefont{G.}~\bibnamefont{Vidal}}, \bibnamefont{and}
  \bibinfo{author}{\bibfnamefont{J.}~\bibnamefont{Cirac}},
  \bibinfo{journal}{Phys.\ Rev. A} \textbf{\bibinfo{volume}{62}},
  \bibinfo{pages}{062314} (\bibinfo{year}{2000}).

\bibitem[{\citenamefont{Brylinski}()}]{Bry}
\bibinfo{author}{\bibfnamefont{J.-L.} \bibnamefont{Brylinski}},
  \eprint{quant-ph/0008031}.

\bibitem[{\citenamefont{Brylinski and Brylinski}()}]{Bry2}
\bibinfo{author}{\bibfnamefont{J.-L.} \bibnamefont{Brylinski}}
  \bibnamefont{and}
  \bibinfo{author}{\bibfnamefont{R.}~\bibnamefont{Brylinski}},
  \eprint{quant-ph/0010101}.

\bibitem[{\citenamefont{Verstraete et~al.}(2002)\citenamefont{Verstraete,
  Dehaene, Moor, and Verschelde}}]{Ver}
\bibinfo{author}{\bibfnamefont{F.}~\bibnamefont{Verstraete}},
  \bibinfo{author}{\bibfnamefont{J.}~\bibnamefont{Dehaene}},
  \bibinfo{author}{\bibfnamefont{B.~D.} \bibnamefont{Moor}}, \bibnamefont{and}
  \bibinfo{author}{\bibfnamefont{H.}~\bibnamefont{Verschelde}},
  \bibinfo{journal}{Phys.\ Rev. A} \textbf{\bibinfo{volume}{65}},
  \bibinfo{pages}{052112} (\bibinfo{year}{2002}).

\bibitem[{\citenamefont{Miyake}()}]{Mi}
\bibinfo{author}{\bibfnamefont{A.}~\bibnamefont{Miyake}},
  \eprint{quant-ph/0206111}.

\bibitem[{\citenamefont{Klyachko}()}]{Klyachko}
\bibinfo{author}{\bibfnamefont{A.~A.} \bibnamefont{Klyachko}},
  \eprint{quant-ph/0206012}.

\bibitem[{\citenamefont{Bennett et~al.}(2001)\citenamefont{Bennett, Popoescu,
  Rohrlich, Smolin, and Thapliyal}}]{Ben}
\bibinfo{author}{\bibfnamefont{C.}~\bibnamefont{Bennett}},
  \bibinfo{author}{\bibfnamefont{S.}~\bibnamefont{Popoescu}},
  \bibinfo{author}{\bibfnamefont{D.}~\bibnamefont{Rohrlich}},
  \bibinfo{author}{\bibfnamefont{J.}~\bibnamefont{Smolin}}, \bibnamefont{and}
  \bibinfo{author}{\bibfnamefont{A.}~\bibnamefont{Thapliyal}},
  \bibinfo{journal}{Phys.\ Rev. A} \textbf{\bibinfo{volume}{63}},
  \bibinfo{pages}{012307} (\bibinfo{year}{2001}).

\bibitem[{\citenamefont{Thibon}(1991)}]{Th91}
\bibinfo{author}{\bibfnamefont{J.-Y.} \bibnamefont{Thibon}},
  \bibinfo{journal}{Internat. J. Alg. Comp.} \textbf{\bibinfo{volume}{2}},
  \bibinfo{pages}{207} (\bibinfo{year}{1991}).

\bibitem[{\citenamefont{Scharf et~al.}(1993)\citenamefont{Scharf, Thibon, and
  Wybourne}}]{STW}
\bibinfo{author}{\bibfnamefont{T.}~\bibnamefont{Scharf}},
  \bibinfo{author}{\bibfnamefont{J.-Y.} \bibnamefont{Thibon}},
  \bibnamefont{and} \bibinfo{author}{\bibfnamefont{B.}~\bibnamefont{Wybourne}},
  \bibinfo{journal}{J. Phys. A: Math. Gen.} \textbf{\bibinfo{volume}{24}},
  \bibinfo{pages}{7461} (\bibinfo{year}{1993}).

\bibitem[{\citenamefont{Segre}(1922)}]{Segre}
\bibinfo{author}{\bibfnamefont{C.}~\bibnamefont{Segre}}, \bibinfo{journal}{Ann.
  di Mat., Ser. III} \textbf{\bibinfo{volume}{29}}, \bibinfo{pages}{105}
  (\bibinfo{year}{1922}).

\bibitem[{\citenamefont{Olver}(1999)}]{Olver}
\bibinfo{author}{\bibfnamefont{P.}~\bibnamefont{Olver}},
  \emph{\bibinfo{title}{Classical Invariant Theory}}
  (\bibinfo{publisher}{Cambridge University Press},
  \bibinfo{address}{Cambridge}, \bibinfo{year}{1999}).

\bibitem[{\citenamefont{Gelfand et~al.}(1994)\citenamefont{Gelfand, Kapranov,
  and Zelevinsky}}]{GKZ}
\bibinfo{author}{\bibfnamefont{I.}~\bibnamefont{Gelfand}},
  \bibinfo{author}{\bibfnamefont{M.}~\bibnamefont{Kapranov}}, \bibnamefont{and}
  \bibinfo{author}{\bibfnamefont{A.}~\bibnamefont{Zelevinsky}},
  \emph{\bibinfo{title}{Discriminants, Resultants, and Multidimensional
  Determinants}} (\bibinfo{publisher}{Birkh{\"a}user},
  \bibinfo{address}{Boston}, \bibinfo{year}{1994}).

\bibitem[{\citenamefont{Schl{\"a}fli}(1852)}]{Schl}
\bibinfo{author}{\bibfnamefont{L.}~\bibnamefont{Schl{\"a}fli}},
  \bibinfo{journal}{Denkschr. der Kaiserl. Akad. der Wiss., math-naturwiss.
  Klasse,} \textbf{\bibinfo{volume}{4}} (\bibinfo{year}{1852}).

\bibitem[{\citenamefont{Paige}(1881)}]{LeP}
\bibinfo{author}{\bibfnamefont{C.~L.} \bibnamefont{Paige}},
  \bibinfo{journal}{Bull. Acad. Roy. Sci. Belgique (3)}
  \textbf{\bibinfo{volume}{2}}, \bibinfo{pages}{40} (\bibinfo{year}{1881}).

\bibitem[{\citenamefont{Schwartz}(1922)}]{Sw}
\bibinfo{author}{\bibfnamefont{E.}~\bibnamefont{Schwartz}},
  \bibinfo{journal}{Math. Z.} \textbf{\bibinfo{volume}{12}},
  \bibinfo{pages}{18} (\bibinfo{year}{1922}).

\bibitem[{\citenamefont{Thrall and Chanler}(1938)}]{TC}
\bibinfo{author}{\bibfnamefont{R.}~\bibnamefont{Thrall}} \bibnamefont{and}
  \bibinfo{author}{\bibfnamefont{J.}~\bibnamefont{Chanler}},
  \bibinfo{journal}{Duke Math. J.} \textbf{\bibinfo{volume}{4}},
  \bibinfo{pages}{678} (\bibinfo{year}{1938}).

\bibitem[{\citenamefont{Chanler}(1939)}]{Chan}
\bibinfo{author}{\bibfnamefont{J.}~\bibnamefont{Chanler}},
  \bibinfo{journal}{Duke Math. J.} \textbf{\bibinfo{volume}{5}},
  \bibinfo{pages}{552} (\bibinfo{year}{1939}).

\bibitem[{\citenamefont{Tapia}()}]{TaP}
\bibinfo{author}{\bibfnamefont{V.}~\bibnamefont{Tapia}},
  \eprint{math-ph/0208010}.

\bibitem[{\citenamefont{Grassl et~al.}()\citenamefont{Grassl, Beth,
  R{\"o}tteler, and Makhlin}}]{GBRM}
\bibinfo{author}{\bibfnamefont{M.}~\bibnamefont{Grassl}},
  \bibinfo{author}{\bibfnamefont{T.}~\bibnamefont{Beth}},
  \bibinfo{author}{\bibfnamefont{M.}~\bibnamefont{R{\"o}tteler}},
  \bibnamefont{and} \bibinfo{author}{\bibfnamefont{Y.}~\bibnamefont{Makhlin}},
  \eprint{Talk at {MSRI}, {N}ovember 19, 2002}.

\bibitem[{\citenamefont{Luque and Thibon}()}]{LT}
\bibinfo{author}{\bibfnamefont{J.-G.} \bibnamefont{Luque}} \bibnamefont{and}
  \bibinfo{author}{\bibfnamefont{J.-Y.} \bibnamefont{Thibon}},
  \eprint{math-ph/02011044}.

\end{thebibliography}

\end{document}